\documentclass[reprint, amsmath, amssymb, aps]{revtex4-2}
\usepackage{graphicx}
\usepackage{dcolumn}
\usepackage{bm}
\usepackage{mathrsfs}
\usepackage{accents}
\usepackage{amsmath}
\newcommand{\ket}[1]{{|#1\rangle}}
\newcommand{\bra}[1]{{\langle#1|}}
\newcommand{\braket}[2]{{\langle#1|#2\rangle}}
\renewcommand{\theequation}{{\rm\arabic{section}.\arabic{equation}}}

\begin{document}

\title{Quantum Kinematics in Terms of Observable Quantities, and \\
the Chirality of Entangled Two-Qubit States}

\author{Daniel F. V. James}
\affiliation{Department of Physics, University of Toronto\\
60, St. George St., Toronto, Ontario, Canada M5S 1A7}
\date{\today}
\begin{abstract}
We consider the kinematics of bi-partite quantum states as determined by observable quantities, in particular the Bloch vectors of the subsystems.  In examining the simplest case of a pair of two-level systems, there is a remarkable connection between the presence of non-classical correlations and the chirality of the two bases generated by the singular value decomposition of the correlation matrix of the Bloch vectors.  We investigate the limits imposed by quantum mechanics of this effect and it relationship with other methods on quantifying the system's non-classical behaviour.
\end{abstract}
\maketitle
\section{Introduction}
Entanglement is the intimate correlation between two or more separated systems peculiar to quantum mechanics \cite{Schrodinger35}.  The past three decades have seen a remarkable growth in interest in this phenomenon, spurred on by the promise of revolutionary applications in communications and information processing \cite{standardQItext}.  For bipartite systems in a pure quantum state, entanglement is well characterized and can be quantified reasonably uncontroversially, for example by calculating the degree of purity or entropy of the reduced density matrix of one of the sub-systems.  However, extending this characterization to mixed states remains problematical.  Even in the simplest non-trivial case, that of a pair of two-level systems (or qubits), a large number of generalizations of the definition of quantum entanglement have been proposed, and none can be unequivocally declared preeminent \cite{RecentEntanglementReview}. As a general rule, these are all based on some notion of information content of the system; For example, Wootters's famous work \cite{Wootters98} provides a compact, reasonably easy-to-evaluate formula for the entanglement of formation of a mixed state of two qubits (see Appendix A for a discussion). The Entanglement of Formation has its maximum if the state is a fully entangled pure state, and is zero if and only if the state is separable (in the sense defined by Werner \cite{Werner89}).  And yet, to quote Wootters: ``there remains a basic question concerning the {\em interpretation} of the entanglement of formation that has not yet been resolved''. The veritable menagerie of related measures of non-classical correlations have since been investigated by other authors \cite{LangCavesShaji11,QAQJ12}; after 25 years, it would be a bold quantum mechanic who would assert the existence of a single unequivocal quantitative metric of quantumness.

Important though quantum information is, we must not lose sight of the fact that it is one possible application of a physical effect, and that quantum mechanics is {\em not} a branch of information theory or computer science.  By fixating on quantum information, which may or may not ultimately lead to a widely deployable technology, we may be overlooking fundamental physical aspects of the phenomenon, as well implicitly marginalizing other possible applications, as yet unknown and unexplored.  Hence a physically motivated, easily calculable operational method in which to define and quantify the notion of quantum coherence remains strongly desirable.

In this paper we approach this problem from such a different perspective.  Heretofore the {\em density operator} $\rho$ has been regarded as the operational quantity used to specify a multi-partite quantum state, and attempts to characterize entanglement all ultimately rest on the evaluation of some function of $\rho$.  However we suggest that this seems a peculiar approach, given the fact that $\rho$ cannot be observed directly, and must instead be deduced via quantum state tomography \cite{JMKW00,Eisert20}, which, even for quite simple systems, involves a burdensome program of data collection and computational reduction.  The observable quantities in quantum mechanics are the correlations between the results of joint measurements carried out upon systems prepared in the appropriate state; an ensemble (in principle infinite, in practice of finite size) of such measurements uniquely specifies the state.  Indeed, Hardy has shown one can derive a consistent quantum theory using such quantities as the operational variables \cite{Hardy01}.  The {\em Bloch vector} provides a mathematically compact, elegant and intuitive way in which to arrange these observable quantities for the simple case of a single two-level system.  This formulation was introduced to describe the dynamics of nuclear spins in nuclear magnetic resonance experiments \cite{Rabi} (although historically similar concepts had been considered much earlier to describe rigid body dynamics \cite{Cayley} and optical polarization \cite{Stokes}).  Bloch vectors have been found to be invaluable in the description of the dynamics of single two-level atomic systems \cite{FVH, AE}, so much so that we should be loath to abandon them when considering two or more such systems. Specifically, here we examine whether the correlations between the Bloch vectors for individual qubits can be used to investigate quantum correlations of a multi-qubit system without first having to deduce the density matrix.  While properties of the density matrix are employed in the derivation of our results, ultimately this results can be calculated without resorting to the tribulation of tomographic state reconstruction.

The paper is organized as follows: first, we consider the properties of the Bloch vector representation of qubit pairs, and introduce the {\em Sinisterness} $\mathcal{S}$ as a quantitive measure of correlation.  We next consider the relationship between this quantity and the density operator, which will be useful in establishing its mathematical properties.  We illustrate our results by calculating the Sinisterness for a number of important categories of quantum states, which serves to illustrate its connection with other types of correlation, in particular the Concurrence. A number of the more involved mathematical derivations are included in the Appendices. A preliminary report of the results presented here was published in \cite{DFVJ14}.

\section{Bloch Vectors and their Correlations}
\subsection{Definitions}
For the sake of this discussion, let us consider the state of each qubit to be specified by a real stochastic three-dimensional vector.  For a single two-level system, the average of this vector is the {\em Bloch vector}, defined as follows:
\begin{equation}
{\bf a}=\langle {\boldsymbol\sigma} \rangle 
\equiv Tr\{\rho^{(A)}{\boldsymbol\sigma}\},
\label{BlochVecDef}
\end{equation}
where ${\boldsymbol\sigma}$ is the vector formed of the Pauli operators, with cartesian components
$\sigma_1=\ket{1}\bra{2}+\ket{2}\bra{1}$,  
$\sigma_2=i\ket{1}\bra{2}-i\ket{2}\bra{1}$ and  
$\sigma_3=\ket{2}\bra{2}-\ket{1}\bra{1}$, ($\ket{1}$ and $\ket{2}$ being the two levels of the system), and $\rho^{(A)}$ is the density operator, which we interpret as a compact means of specifying the probability densities of the random variables associated with our system.  For a single isolated qubit these probabilities are all completely specified by the Bloch vector; specifically, the density operator is related to the Bloch vector by the formula:
\begin{equation}
\rho^{(A)} = \frac{1}{2}
\left({\rm I}+{\bf a}\cdot{\boldsymbol\sigma}\right), 
\end{equation}
where ${\rm I}=\ket{2}\bra{2}+\ket{1}\bra{1}$ is the identity operator.

For the bipartite system, besides the average values of the two Bloch vectors for the individual qubits, ${\bf a}$ and ${\bf b}$, we also consider the  {\em cross-correlation tensor} of the two spins, defined by
\begin{equation}
c_{ij} =\frac{1}{4} \langle (\sigma_i-a_i{\rm I})\otimes(\sigma_j-b_j{\rm I})\rangle,
\end{equation}
where we have introduced the tensor product $\otimes$ to combine the two spaces of individual systems, with ordering implying on which sub-space operators act.  The matrix $c_{ij}$ is real, but not necessarily symmetric; its nine components, together with the six components of ${\bf a}$ and ${\bf b}$ completely specify the probability distribution of the bipartite system:
\begin{equation}
\rho^{(AB)} =
\frac{1}{2} \left({\rm I}+{\bf a}\cdot{\boldsymbol\sigma}\right)
\otimes
\frac{1}{2} \left({\rm I}+{\bf b}\cdot{\boldsymbol\sigma}\right)
+
\sum_{i,j=1}^3 c_{ij}\sigma_i\otimes\sigma_j.
\label{FanoForm}
\end{equation}
Decompositions of this kind were first considered by Fano some time ago \cite{Fano} and have found considerable utility in the theory of entanglement and in quantum state tomography \cite{JMKW00}. In particular, the geometric properties of a related correlation tensor and its connection to quantum coherence have been discussed by Englert and Metwally \cite{Englert02}. 

\subsection{Properties of the Coherence Matrix}
If all the components of $c_{ij}$ are zero, the state is factorable, and the two sub-systems are completely uncorrelated.  If that is not the case, we are immediately confronted with the issue of how to interpret these correlations.  To begin this task, we invoke the principle that coherence properties must be independent of any local change of basis.  For Bloch vectors, a unitary operation is equivalent to a transformation of three-dimensional coordinate axes introduced implicitly by the definition of $\sigma_1$,  $\sigma_2$ and $\sigma_3$. Thus it behoves us to choose some particular set of axes which will simplify the system.  Particularly useful sets of axes is provided by the {\em singular value decomposition} \cite{SVDref}, of the matrix $c_{ij}$, which we shall write as follows:  
\begin{equation}
c_{ij} = \sum_{p,q=1}^{3} {\rm U}_{ip}\Sigma_{pq}{\rm V}^T_{qj},
\label{SVD}
\end{equation}
where ${\rm U}$ and ${\rm V}$ are orthogonal matrices, and $\Sigma$ is a real, non-negative definite diagonal matrix. We shall denote the singular values (i.e. the diagonal elements of $\Sigma$) as $s_p$ $(p=1,2,3)$, thus eq.(\ref{FanoForm}) may be re-written as
\begin{equation}
\rho^{(AB)} =
\frac{1}{2}\left({\rm I}+ {\bf a}^\prime\cdot{\boldsymbol\alpha}\right)
\otimes
\frac{1}{2}\left({\rm I}+ {\bf b}^\prime\cdot{\boldsymbol\beta}\right)
+\sum_{p=1}^3 s_p\alpha_p\otimes\beta_p,
\end{equation}
where  $a_i^\prime= \sum_j{\rm U}^T_{ij}a_j$,  $b_i^\prime= \sum_i{\rm V}^T_{ij}b_j$, 
$\alpha_i= \sum_j{\rm U}^T_{ij}\sigma_j$ and  $\beta_i= \sum_j{\rm V}^T_{ij}\sigma_j$.

However this is {\em not} simply a special case of eq.(\ref{FanoForm}) written with new coordinate axes so that the cross-correlation term is diagonal.  
While the matrices ${\rm U}$ and ${\rm V}$ are real orthogonal matrices in three dimensions, they are not necessarily {\em proper} rotations, implying that the two sets of system-dependent operators 
$\{\alpha_1, \alpha_2, \alpha_3\}$ and $\{\beta_1, \beta_2, \beta_3\}$ do not necessarily obey the standard commutation relations for Pauli operators. 
Thus, for example,
\begin{eqnarray}
\left[\alpha_p,\alpha_q\right]&=&\sum_{i,j=1}^3 {\rm U}^T_{pi}{\rm U}^T_{qj} \left[\sigma_i,\sigma_j\right]\nonumber\\
&=&\sum_{i,j=1}^3 {\rm U}^T_{pi}{\rm U}^T_{qj} \left( 2i \sum_{k=1}^3\epsilon_{ijk}\sigma_k\right)\nonumber\\
&=&2i\sum_{i,j,k,r=1}^3 {\rm U}^T_{pi}{\rm U}^T_{qj} \epsilon_{ijk} {\rm U}_{kr}\alpha_r\nonumber\\
&=& 2i\sum_{r=1}^3\epsilon^\prime_{pqr}\alpha_r
\label{comrelalpha}
\end{eqnarray}
where $\epsilon_{ijk}$ is the usual Levi-Civita tensor and $\epsilon^\prime_{pqr}=\sum_{i,j,k=1}^3 \epsilon_{ijk} {\rm U}^T_{pi}{\rm U}^T_{qj} {\rm U}^T_{rk} $.  
Like $\epsilon_{ijk}$, the transformed tensor $\epsilon^\prime_{pqr}$ is completely anti-symmetric, e.g.
\begin{eqnarray}
\epsilon^\prime_{prq}&=&\sum_{i,j,k=1}^3 \epsilon_{ijk} {\rm U}^T_{pi}{\rm U}^T_{rj} {\rm U}^T_{qk}
\nonumber\\
&=&\sum_{i,j,k=1}^3 \epsilon_{ikj} {\rm U}^T_{pi}{\rm U}^T_{rk} {\rm U}^T_{qj}
\nonumber\\
&=&-\sum_{i,j,k=1}^3 \epsilon_{ijk} {\rm U}^T_{pi}{\rm U}^T_{qj}{\rm U}^T_{rk} 
\nonumber\\
&=&-\epsilon^\prime_{pqr}.
\end{eqnarray}
Similar arguments may be applied to interchanging any two indices. Thus it follows that $\epsilon^\prime_{pqr}=(\epsilon^\prime_{123})\,\epsilon_{pqr}$, where $\epsilon^\prime_{123} =\mbox{Det}\{{\rm U}^T\}$. Since $\mbox{Det}\{{\rm U}^T\}=\mbox{Det}\{{\rm U}\}$ it follows
\begin{equation}
\epsilon^\prime_{pqr} = \mbox{Det}\{{\rm U}\} \epsilon_{pqr}. \label{epsilontilde}
\end{equation}
Further, since $\mbox{Det}\{{\rm U} {\rm U}^T\}=\mbox{Det}\{{\rm I}\}=1$, it follows that $\mbox{Det}\{{\rm U}\} =\pm 1$; 
When $\mbox{Det}\{{\rm U}\} =1$, the transformation ${\rm U}$ is a proper rotation, under which the chirality of the coordinate system and commutation relation is not changed; however, when $\mbox{Det}\{{\rm U}\} =-1$ the rotation is improper and a right-handed set of basis vectors is transformed into a left-handed set.  
Thus the commutation relation eq.(\ref{comrelalpha}) reduces to
\begin{equation}
\left[\alpha_p,\alpha_q\right] = 2i \mbox{Det}\{{\rm U}\}\sum_{r=1}^3\epsilon_{pqr}\alpha_r.
\end{equation}
Similar remarks apply to $\beta_p$ and ${\rm V}$.  

If both ${\rm U}$ and ${\rm V}$ are improper rotations, it is straightforward to interchange the labels of any two of the singular values $s_p$ to render them into proper rotations. However, if one is proper and the other one improper, this is not possible: swapping the label  of $s_p$ to convert the the improper matrix to be proper will have the reverse effect on the other transformation.

It is a remarkable fact that, {\em for all entangled states, one of these transforms is proper and the other improper}; or geometrically speaking, the two diagonalized coordinate frames for the two qubits have opposite chirality.  Following Dr. Joseph Altepeter, who seems to have been the first to point out a connection between chirality of Bloch vector representations and entanglement \cite{AJMK}, we refer to states with this chiral property as  {\em sinisterness} (from the Latin word for left-handed).  

The determinant of the $3\times 3$ matrix $c_{ij}$ is related to the singular value decomposition in a useful way.  The determinant is given by the usual expression
\begin{equation}
\mathcal{S}\equiv\mbox{Det}\{c\}=\frac{1}{6}\sum_{i,j,k,l,m,n=1}^3\epsilon_{ijk}\epsilon_{lmn}c_{il}c_{jm}c_{kn}.
\label{detdef}
\end{equation}
Substituting eqs.(\ref{SVD}) and (\ref{epsilontilde}) , we find that 
\begin{eqnarray}
\mathcal{S} &=&
\frac{1}{6}\sum_{i,j,k,l,m,n=1}^3\epsilon_{ijk}\epsilon_{lmn}
\nonumber\\
&&\times\sum_{p=1}^3U^T_{pi}s_pV^T_{pl}\sum_{q=1}^3U^T_{qj}s_pV^T_{qm}
\sum_{r=1}^3U^T_{rk}s_pV^T_{rn}\nonumber\\
&=&\frac{1}{6}\mbox{Det}\{{\rm U}\}\mbox{Det}\{{\rm V}\} \sum_{p,q,r=1}^3 (\epsilon_{pqr})^2 s_p s_q s_r\nonumber\\
&=& \mbox{Det}\{{\rm U}\}\mbox{Det}\{{\rm V}\} s_1 s_2 s_3.
\end{eqnarray}
The singular values $s_k$ are all positive, while determinants are either $+1$ for a proper rotation, or $-1$ for an improper rotation.  Thus a simple way to determine whether or not a state is sinister is to calculate the determinant of its correlation matrix $c_{ij}$: if it is negative, the state is sinister.  Indeed, as we shall see, the value of this determinant in fact reveals some quantitative features of the state, so much so that we suggest it may be useful considering it as an alternative way to quantify the quantumness of a state.  Accordingly, we define the {\em Sinisterness} of a state to be given by the formula
\begin{eqnarray}
\mathcal{S}&=&\mbox{Det}\{c_{ij}\}\\
&=& \mbox{Det}\{\langle (\sigma_i-a_i{\rm I})\otimes(\sigma_j-b_j{\rm I})\rangle/4\}
\label{Sdef}
\end{eqnarray}
One important property is readily apparent: since the value of the determinant is invariant under proper rotations, the Sinisterness is invariant under local unitaries, an important feature usually required of any measure of quantum correlation. 

\section{Relationships between $\mathcal{S}$ and the density operator $\rho$}
While $\mathcal{S}$ is designed to be easily calculated from observable correlations without the necessity of quantum state estimation, in order to explore its theoretical implications it is useful to relate $\mathcal{S}$ to the density operator $\rho$ of the underlying state.  

One useful identity in calculating the $\mathcal{S}$ can be found by labeling the identity matrix as $\sigma_0$ and considering the following $4\times4$ matrix 
\begin{equation}
\Gamma_{\mu,\nu}= \langle \sigma_{\mu}\otimes \sigma_{\mu}\rangle\,\,\, (\mu,\nu=0,1,2,3),
\label{gammadef}
\end{equation}
we find that $\Gamma_{0,0}=1$, $\Gamma_{i,0}=a_i$, $\Gamma_{0,j}=b_j$ and
$\Gamma_{i,j}=c_{ij}+a_ib_j$.  As a side note, the matrix $\Gamma$ can be used to find the purity of the quantum state without the need for tomography, since $\mbox{Tr}\{\Gamma^T\Gamma\}=\mbox{Tr}\{\rho^2\}$. 

Introducing two matrices
\begin{equation}
{\rm L}=\left(
\begin{array}{cccc}
1&0&0&0\\
-a_1&1&0&0\\
-a_2&0&1&0\\
-a_3&0&0&1
\end{array}
\right)\,\,\,\mbox{and}\,\,\,
{\rm R}=\left(
\begin{array}{cccc}
1&-b_1&-b_2&-b_3\\
0&1&0&0\\
0&0&1&0\\
0&0&0&1
\end{array}
\right),\end{equation}
both of which have unit determinant, we find
\begin{equation}
{\rm L} \Gamma {\rm R}=
\left(
\begin{array}{cccc}
1&0&0&0\\
0&c_{xx}&c_{xy}&c_{xz}\\
0&c_{yx}&c_{yy}&c_{yz}\\
0&c_{zx}&c_{zy}&c_{zz}
\end{array}
\right),
\end{equation}
and hence, since the determinant of a product of matrices is the product of the determinants,
\begin{equation}
\mbox{Det}\{\Gamma\}=\mbox{Det}\{c\}=\mathcal{S}.
\end{equation}

Equation (\ref{gammadef}) implies that the elements of the coherence matrix $\Gamma$ are all linear combinations of the elements of the density matrix. One can show that 
\begin{equation}
\Gamma = \frac{1}{2} {\rm W}\mathscr{G}{\rm W}^T
\label{GamfroG}
\end{equation}
where the superscript $T$ denotes the transpose ({\em not} the Hermitian transpose) of the matrix and the matrix ${\rm W}$ is given
\begin{equation}
{\rm W}=
\frac{1}{\sqrt{2}}\left(
\begin{array}{cccc}
1&0&0&1\\
0&1&1&0\\
0&i&-i&0\\
1&0&0&-1
\end{array}
\right).
\end{equation}

The matrix $\mathscr{G}$ appearing in eq.(\ref{GamfroG}) is formed from a transposition of certain elements of the density matrix, viz.,
\begin{eqnarray}
\mathscr{G}&=&
\sum_{m,n=0}^1 ({\rm I}\otimes \ket{m}\bra{n})\rho(\ket{m}\bra{n}\otimes{\rm I})
\nonumber\\
&=&
\left(
\begin{array}{cccc}
\rho_{00,00}&\rho_{00,01}&\rho_{01,00}&\rho_{01,01}\\
\rho_{00,10}&\rho_{00,11}&\rho_{01,10}&\rho_{01,11}\\
\rho_{10,00}&\rho_{10,01}&\rho_{11,00}&\rho_{11,01}\\
\rho_{10,10}&\rho_{10,11}&\rho_{11,10}&\rho_{11,11}\\
\end{array}
\right),\label{Gdef}
\end{eqnarray}
where $\rho_{ij,kl}=\langle ij |\rho_{AB}| kl\rangle\,\,\, (i,j,k,l = 0,1)$ are the matrix elements of the density operator $\rho$ in the standard 2-qubit computational  basis (i.e. $\ket{ij}_{AB}=\ket{i}_A\otimes\ket{j}_B$) \.  
Suggestively, $\mathscr{G}$ can also be written in terms of the partial transpose operation \cite{PeresPPT} as follows:
\begin{eqnarray}
\mathscr{G}
&=&
SWAP\cdot (SWAP\cdot\rho)^{T_A}\nonumber\\
&=&
(\rho\cdot SWAP)^{T_B}\cdot SWAP,
\end{eqnarray}
where ${T_A}$ and ${T_B}$ denote partial transpose with respect to the first and second qubit, respectively, and $SWAP$ is the standard SWAP-gate (see ref. \cite{standardQItext}, Sec.1.3.4, p.23).

Taking the Determinant of $\Gamma$ as defined by eq.(\ref{GamfroG}), and using $\mbox{Det}\{{\rm W}\}=\mbox{Det}\{{\rm W^T}\}=i$ we obtain the rather simple expression for the Sinisterness
\begin{equation}
\mathcal{S}=\mbox{Det}\{\Gamma\}=-16\mbox{Det}\{\mathscr{G}\}.
\label{SinisDef}
\end{equation}
This expression is most useful in calculating $\mathcal{S}$ in situations when the density operator is known, e.g. when one is investigating the properties of some model state or a simulation of dynamics using a solution of a master equation.

\section{Some Examples}
\subsection{Pure States}
A pure state can be written in the standard computational basis introduced above, as follows:
\begin{equation}
\ket{\psi}=\alpha \ket{0}+\beta\ket{1}+\gamma\ket{2}+\delta\ket{3}
\end{equation}
where $|\alpha|^2+|\beta|^2+|\gamma|^2+|\delta|^2=1$.
The $4\times4$ matrix $\mathscr{G}$, defined in eq.(\ref{Gdef}), is given by:
\begin{equation}
\mathscr{G} =
\left(
\begin{array}{cccc}
|\alpha|^2 &		\alpha\beta^*&		\beta\alpha^*&		|\beta|^2\\
\alpha\gamma^*&	\alpha\delta^*& 		\beta\gamma^*& 	\beta\delta^*\\
 \gamma\alpha^*&	\gamma\beta^*&	\delta\alpha^*&		\beta\delta^*\\
 |\gamma|^2&		\gamma\delta^*& 	\delta\gamma^*&	|\delta|^2
\end{array}
\right).
\end{equation}
The determinant can be found by some straightforward if tedious algebraic manipulation to be:
\begin{equation}
\mathcal{S} =-16\mbox{Det}\{\mathscr{G}\}=-(2 |\alpha\delta-\beta\gamma|)^4.
\end{equation}
The quantity $\mathcal{C}=2 |\alpha\delta-\beta\gamma|$ is called the {\em Concurrence} and is well known in the study of entangled systems (see Appendix D). For a separable (i.e. factorizible) state $ac\ket{00}+ad\ket{01}+bc\ket{10}+bd\ket{11}$ (where $a,b,c$ and $d$ are complex numbers and $|a|^2+|b|^2=|c|^2+|d|^2=1$), we find by direct substitution $\alpha\delta=\beta\gamma=abcd$ and hence $\mathcal{C}=0$.  Thus for pure states
\begin{equation}
\mathcal{S}_P=-\mathcal{C}_P^4.
\end{equation}

\subsection{classically correlated mixed states}
Moving on to mixed states, and following the hierarchy of quantum correlations introduced in ref. \cite{MPSVW},
{\em Classical} states are states all of whose eigenstates are separable, and thus the density operator can be rendered diagonal by the action of local unitaries acting on the two sub-systems independently. Since correlations are independent of local unitaries, we can without loss of generality represent our state in this diagonal form.  If the diagonal elements are $\{p_{0},p_{1},p_{2},p_{3}\}$, then the components of the $\mathscr{G}$ matrix are
\begin{equation}
\mathscr{G}_{C} =
\left(
\begin{array}{cccc}
p_1& 0&0&p_2\\
0&0& 0& 0\\
0&0& 0& 0\\
p_3&0& 0&p_4
\end{array}
\right).
\end{equation}
Since both two rows and two columns are filled with zeroes, this matrix has determinant zero, and so we find $\mathcal{S}_C=0$.

\subsection{separable mixed states}
Separable states are states without entanglement, but which nevertheless may exhibit some other non-classical correlations.  Following Werner \cite{Werner89}, these may be written as a mixture of factorable pure states, as follows:
\begin{equation}
\rho^{(AB)}_{separable} =\sum_{n=1}^N p_n \ket{\Psi_n}\bra{\Psi_n} \otimes \ket{\Phi_n}\bra{\Phi_n},
\end{equation}
where $0\le p_i\le 1$ and $\sum_{i=1}^Np_i=1$ (such weighted sums are known as {\em convex hulls}). 

There is, in general, an infinite number of such a decompositions for any separable state; however there exists some optimal decomposition for which the number of terms has the smallest possible value, called the {\em optimal ensemble cardinality}, ${\cal L}$, of the state \cite{Divincenzo00}. 
One might conjecture the value of the optimal ensemble cardinality ${\cal L}$ by counting independent parameters: a pair of $d$-level quantum systems has  Hilbert space of dimension $\mathcal{N}=d^2$; a density matrix in an $\mathcal{N}$-dimensional space has $\mathcal{N}^2-1=d^4-1$ independent real parameters.  A normalized, pure state of a $d$-level system with an arbitrary global phase has $2(d-1)$ real parameters, hence the Werner decomposition of a separable state of two $d$-level systems with $N$ terms in the convex hull has $(4(d-1)+1)N-1$ real parameters, when we take into account the weighting factors and the normalization.  Thus, na\"ively, one might expect an arbitrary separable state should have an optimal ensemble cardinality ${\cal L} \le \lceil d^4/(4d-3)\rceil $. In the case $d=2$, this gives ${\cal L} \le 4$; Indeed Wootters has demonstrated how to construct a Werner decomposition for a separable pair of two-level systems with at most four terms (see \cite{Wootters98}, eq.(23)), confirming that ${\cal L}\le 4$ in this case.   

Let us thus write the ``optimal'' Werner decomposition as follows:
\begin{equation}
\rho^{(AB)}_{separable} = 
\sum_{n=1}^4 p_n
\ket{\psi_n}\bra{\psi_n}
\otimes 
\ket{\phi_n}\bra{\phi_n},
\end{equation}
where $\sum_{n=1}^{4} p_n =1$, and $p_n\ge0$ (and we have included the case that ${\cal L}<4$ by allowing the value of $p_n$ to be zero).  Without loss of generality, we can assume that $p_1\ge p_2\ge p_3\ge p_4$.  Defining the unit vectors ${\bf a}_{n}=\bra{\psi_n}{\boldsymbol\sigma}\ket{\psi_n}$ and 
${\bf b}_{n}=\bra{\phi_n}{\boldsymbol\sigma}\ket{\phi_n}$, we find
\begin{equation}
c_{ij}=\sum_{n=1}^4p_n (a_{n,i} -\bar{a}_{i})(b_{n,j}-\bar{b}_{j}),
\end{equation}
where $\bar{\bf a}=\sum_{n=1}^4p_n\bar{\bf a}_n$ and $\bar{\bf b}=\sum_{n=1}^4p_n\bar{\bf b}_n$.  In Appendix A it is shown that the determinant of this matrix (and thus the Sinisterness) can be written in the following relatively compact form
\begin{equation}
\mathcal{S}_{sep}= 36\,p_1p_2p_3p_4\,
\mathcal{V} ({\bf a}_1, {\bf a}_2, {\bf a}_3, {\bf a}_4)
\mathcal{V} ({\bf b}_1, {\bf b}_2, {\bf b}_3, {\bf b}_4),
\label{littlesum3}
\end{equation}
where $\mathcal{V} ({\bf a}_1, {\bf a}_2, {\bf a}_3, {\bf a}_4)=\left[({\bf a}_1-{\bf a}_2)\times ({\bf a}_2-{\bf a}_3)\right]\cdot ({\bf a}_3-{\bf a}_4)/6$ is an antisymmetric quadruple vector product whose magnitude is equal to the volume of the pyramid whose vertices have the position vectors ${\bf a}_1$, ${\bf a}_2$, ${\bf a}_3$ and ${\bf a}_4$.

Some useful conclusions may be drawn from eq.(\ref{littlesum3}).  Firstly, if optimal ensemble cardinality $\mathcal{L}$ of the separable state is $3$ or less, then by definition, $p_4=0$ and we immediately find that $\mathcal{S}=0$.  Further, since ${\bf a}_n$ and ${\bf b}_n$ are both Bloch vectors corresponding to pure states, they are unit vectors.  As is shown in Appendix B, the maximum volume for a pyramid whose vertices lie on the unit sphere is $8/9\sqrt{3}$. Further, the maximum possible product $p_1p_2p_3p_4$ occurs when $p_1=p_2=p_3=p_4=1/4$.  Thus we find that: 
\begin{equation}
-1/27 \le \mathcal{S}_{sep} \le 1/27
 \end{equation}
with the immediate implication that if $\mathcal{S}<-1/27$ the state must be entangled.  Note that $1/27$ is the upper limit on $\mathcal{S}$ for arbitrary states.
  
\subsection{Mixed Entangled States}
\subsubsection{Werner states}
Turning our attention now to entangled mixed states, consider a simple model for which analytic expressions exist: the Werner state \cite{Werner89}, defined by
\begin{equation}
\rho_W(\epsilon)=\frac{(1-\epsilon)}{4}{\rm I}+\epsilon\Pi,
\end{equation}
where $\epsilon$ is a real number in the range $[-1/3,1]$ and 
$\Pi = \ket{\varphi_{ME}}\bra{\varphi_{ME}}$ is the projector for a maximally entangled state; to simplify the analysis, we choose $\ket{\varphi_{ME}}=\frac{1}{\sqrt{2}}(\ket{00}+\ket{11})$, where we have used the notation for two-qubit states introduced above, following eq.(\ref{Gdef}). The Bloch vectors for a Werner state are $\bf{a}=\bf{b}=0$ and $c_{ij}$ is diagonal with $c_{11}=-c_{22}=c_{33}=\epsilon$, and hence $\mathcal{S}=-\epsilon^3$.

The Werner state has an analytic expression for the Concurrence, and is thus a particularly useful model for comparisons.  The matrix ${\mathscr R}$ defined by equation eq.(\ref{Rdef}) has the form
\begin{equation}
{\mathscr R}_W =\frac{(1-\epsilon)^2}{16}{\rm I}+\frac{\epsilon(\epsilon+1)}{2}\Pi,
\label{RWdef}
\end{equation}
which is Hermitian, which simplifies the analysis considerably.  The eigenvalues are $[(3\epsilon+1)/4]^2$ (corresponding to the eigenvector $\ket{\varphi_{ME}}$) and $[(1-\epsilon+)/4]^2$ (which is triply degenerate, it's eigenvectors being orthogonal to $\ket{\varphi_{ME}}$).  Thus using eq.(\ref{ConcDef2}) we find that the Concurrence is given by the expression
\begin{equation}
\mathcal{C}_W = \mbox{max}\{(3\epsilon-1)/2,0\}.
\end{equation}
Thus, for Werner states with non-zero Concurrence, the Sinisterness $\mathcal{S}$ is simply related to the Concurrence $\mathcal{C}$ by the formula 
\begin{equation}
\mathcal{S} = -\left(\frac{2 \mathcal{C}+1}{3}\right)^3,
\end{equation}
and $\mathcal{S}=-1/27$ when $\epsilon=1/3$, i.e. the largest value of $\epsilon$ for which the Concurrence is zero.

 \subsubsection{X states}
A slightly more general analytic model of mixed, partially entangled states are the so-called ``X-states'', defined by the following density matrix (in the computational basis):
 \begin{equation}
 \rho_X=
 \left(
 \begin{array}{cccc}
 q&0&0&v\\
 0&r&u&0\\
 0&u&s&0\\
v&0&0&t\\
 \end{array}
 \right),
 \label{xstatedef}
 \end{equation}
where $q$, $r$, $s$, $t$, $u$ and $v$ are real positive parameters such that 
$q+r+s+t=1$, $\sqrt{r s}\ge u$ and $\sqrt{qt}\ge v$ (note that a similar model with complex off-diagonal terms can be transformed to a state of the form given by eq.(\ref{xstatedef}) by local transformations, and hence it will have similar coherence properties).  For X states, the Concurrence is $\mathcal{C}_X=2 {\rm Max}\{u-\sqrt{q t},v-\sqrt{r s},0\}$.  Thus if either of the conditions $u>\sqrt{q t}$ or $v>\sqrt{r s}$ are satisfied, then the state is entangled. 

Using eqs.(\ref{Gdef}) and (\ref{SinisDef}), the Sinisterness of an X-state is $\mathcal{S}_X = -16 \mbox{Det}\{\mathscr{G}_X\}$ where
\begin{equation}
\mathscr{G}_X=\left(
\begin{array}{cccc}
 q&0&0&r\\
 0&v&u&0\\
 0&u&v&0\\
s&0&0&t\\
 \end{array}
\right),
\end{equation}
and the determinant  can be straightforwardly calculated to give $\mathcal{S}_X=-16 \left(v^2-u^2\right) (q t-r s)$. 
Suppose that $u>\sqrt{qt}$, so the state is entangled; since $\sqrt{rs}\ge u > \sqrt{qt} \ge v$, we immediately see that $\mathcal{S}<0$; similarly $\mathcal{S}<0$ if $v>\sqrt{rs}$.  In other words, if $\rho_X$ is entangled, $\mathcal{S}<0$.

 \subsubsection{Arbitrary Mixed Entangled States}
For arbitrary mixed entangled states, which do not admit a description in terms of either Werner States, X states or some other analytic model, we have yet to find a general analytic relationship  between Sinisterness and Concurrence or any other quantum correlation.  We can however investigate the relationship by numerical experimentation.  Density matrices must be Hermitian, unit trace and non-negative definite; such matrices can be generated numerically quite simply by means of a random Cholesky matrix decomposition: i.e. $\rho_{random}=T^\dagger T/\mbox{Tr}\{T^\dagger T\}$, where $T$ is a 4x4 matrix with complex elements whose real and imaginary parts are uniformly distributed between 0 and 1. Given a numerical density matrix, the Sinisterness and the Concurrence can be easily be calculated and the results plotted (see Fig.1).  The results are quite suggestive: in particular, for {\em all} entangled states in our sample we found that the Sinisterness of a particular state, $\mathcal{S}(\rho)$ is constrained by the value of the Concurrence $\mathcal{C}(\rho)$ by the following inequality:
\begin{equation}
-\mathcal{C}(\rho)^4 \ge \mathcal{S}(\rho) \ge -\left(\frac {2\,\mathcal{C}(\rho)+1}{3}\right)^3.
\end{equation}
The upper limit corresponds to the pure state case, the lower limit to the Werner states.

\begin{figure}[h]
\centerline{\includegraphics[width=8cm]{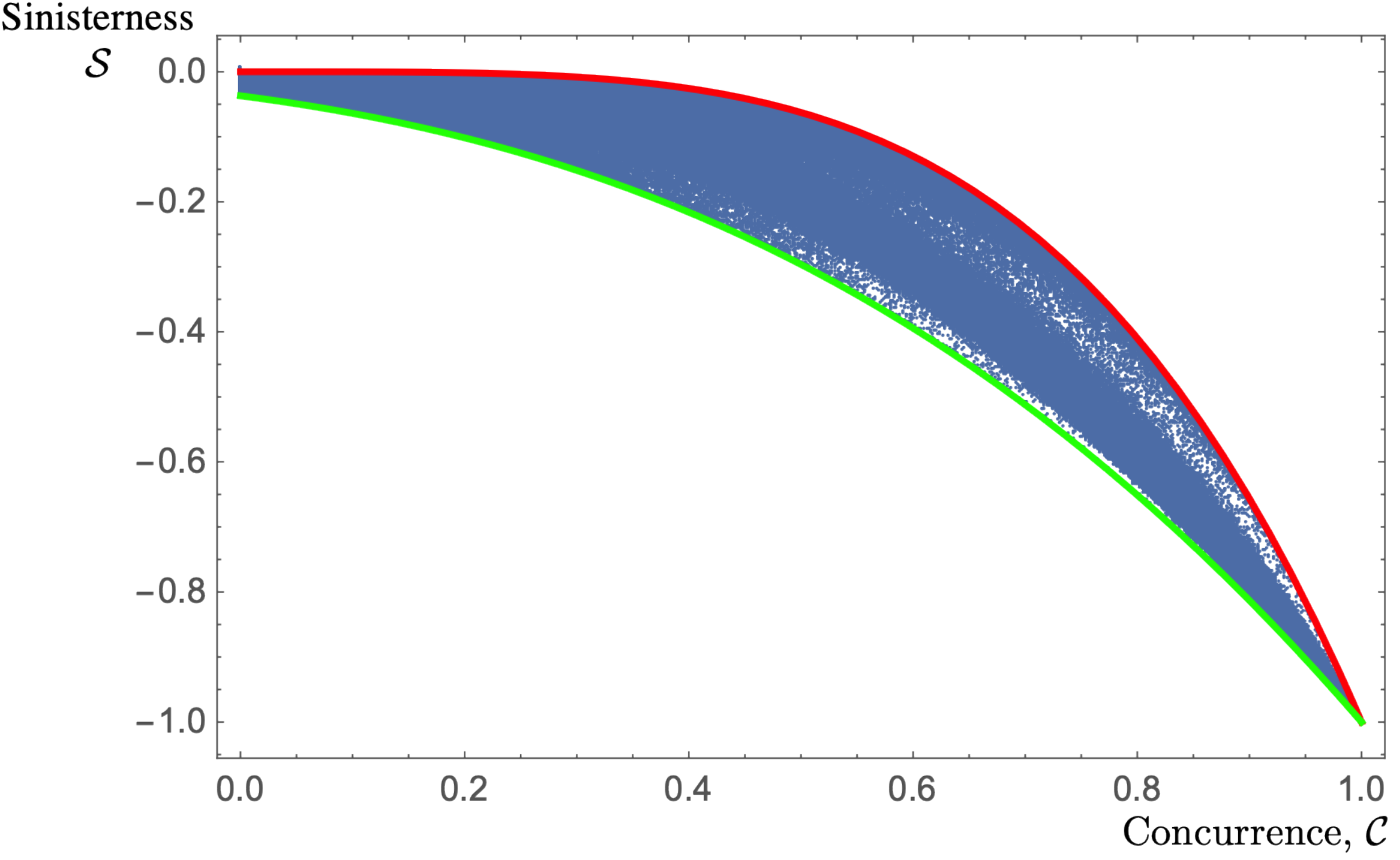}}
\caption{A plot of Concurrence $\mathcal{C}$ (x-axis) and Sinisterness $\mathcal{S}$ for 966,391 randomly selected density matrices. The lines corresponding to pure states ($\mathcal{S}_{max}=-\mathcal{C}^4$) (upper curve, red) and Werner states  ($\mathcal{S}_{min}=-[(2\,\mathcal{C}+1)/3]^3$) (lower curve, green) are also shown; all states in the sample fell between these two extremes. The sample was biased so as to favour states near either the pure or Werner states: this accounts for the two sparse bands in the figure.   For {\em all} states in our sample, the Concurrence was greater than zero if the Sinisterness was negative, supporting the conjecture that {\em all entangled states are sinister}.  Some 17.5\% (169,368 out of 966,391) of the sample were unentangled (i.e. $\mathcal{C}=0$); for these states the Sinisterness $\mathcal{S}$ always fell in the range $1/27 \ge \mathcal{S} \ge-1/27$, as expected.}
\end{figure}

We can provide some slightly more rigorous indication that the Werner States do in fact represent the lower bound on the value of $\mathcal{S}$.  Consider an infinitesimal variation of a Werner State, i.e. $\rho_W\rightarrow (1-\lambda)\rho_W+\lambda\rho^\prime$, where $\lambda \rightarrow 0^+$.  We can obtain simple expressions for the corresponding variations of both Concurrence $\mathcal{C}$, eq.(\ref{deltaCW}), and Sinisterness $\mathcal{S}$, eq.(\ref{deltaSW}), viz:
\begin{eqnarray}
\delta\mathcal{C}_W&=&
2 \lambda\left(\mbox{Tr}\{\Pi\rho^\prime\}-\frac{(3\epsilon+1)}{4}\right) + \rm{O}[\lambda^2],\\
\delta\mathcal{S}_W &=& \mathcal{S}_W \frac{4}{\epsilon}\left(\mbox{Tr}\{\Pi\rho^\prime\}-\frac{(3\epsilon+1)}{4}\right)+ \rm{O}[\lambda^2].
\end{eqnarray}
If we constrain the variation $\rho^\prime$ so that the Concurrence is not changing, i.e. we will be moving only in the vertical direction in Fig.1, then we find that $\delta\mathcal{S}_W=0$, implying this is indeed a stationary point, as one would expect of an extremum.  Attempts to perform a similar analysis of the pure state case have unfortunately met with frustration, due to the complications inherent in bi-orthogonal perturbation in degenerate null subspaces.

\section{Summary and Conclusion}
To summarize our findings about the Sinisterness $\mathcal{S}$ defined by eq.(\ref{Sdef}), it is a quantity which can be determined directly from experimental data, without the need for tomographic reconstruction of the density matrix, and has many of the desirable properties of a measure of quantum coherence.  It is related to the density matrix in a very simple way, eq.(\ref{SinisDef}) which facilitates the theoretical instigation of its properties.  In particular, for {\em arbitrary} states $1/27\ge\mathcal{S}\ge-1$; for classically correlated states, and separable states with an optimal ensemble cardinality $\mathcal{L}$ of $3$ or less  $\mathcal{S}=0$; for separable states with $\mathcal{L}=4$, $1/27\ge\mathcal{S}\ge-1/27$; and for all entangled states its conjectured that $\mathcal{S}<0$, with some strong numerical evidence in support. It is weakly related to the Concurrence $\mathcal{C}$, with the simple functions of $\mathcal{C}$ providing the upper and lower limits on the value of $\mathcal{S}$.

\begin{figure}[h]
\centerline{\includegraphics[width=8cm]{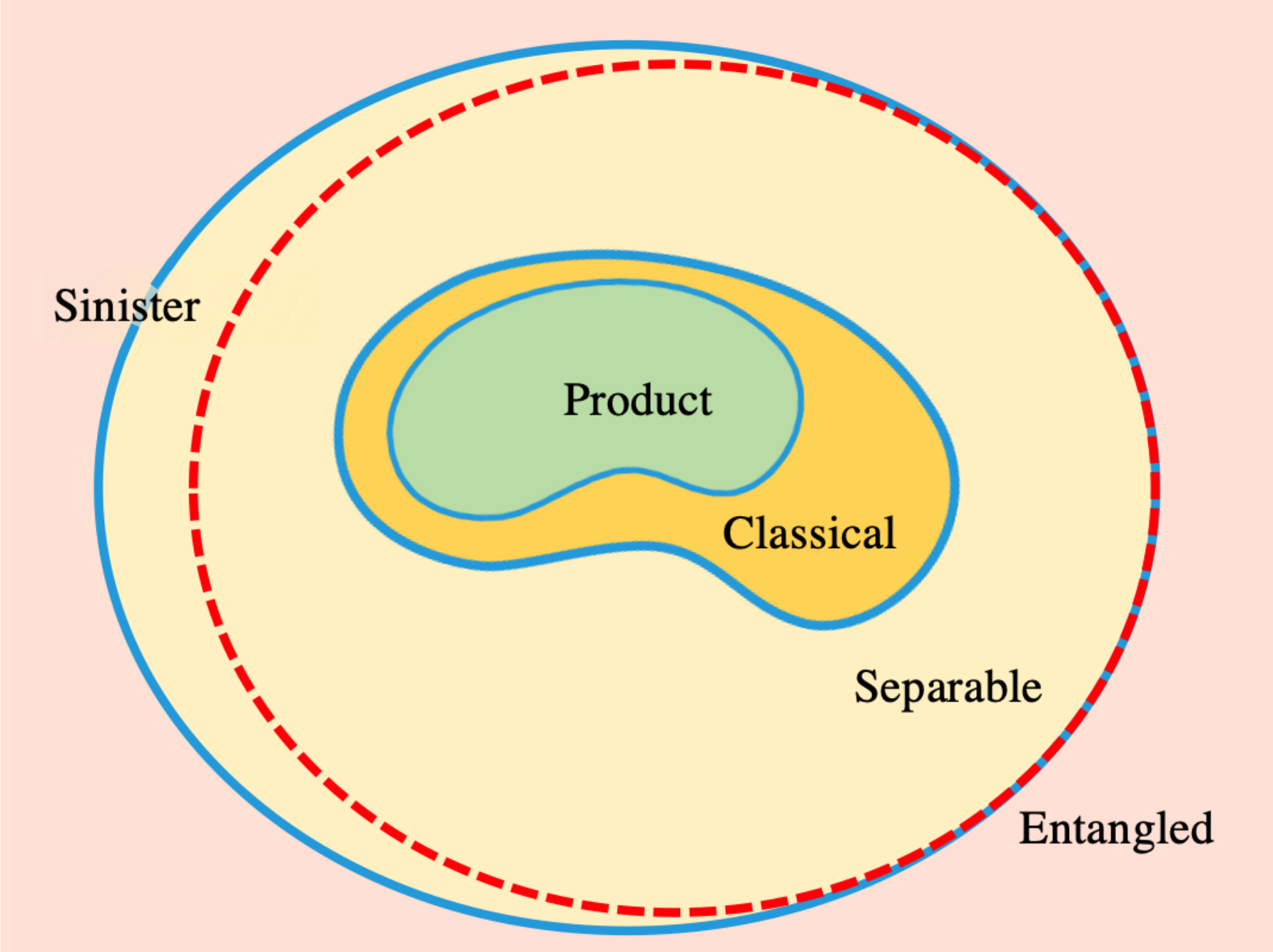}}
\caption{Schematic showing the hierarchy of quantum correlations, including Sinisterness.  The standard hierarchy \cite{MPSVW} of Entangled, Separable, Classically Correlated and Product States are shown with solid blue boundaries.  The domain of Sinister states is shown with a dashed red boundary: all entangled states, and some separable states are sinister, but classically correlated and product states are not.}
\label{fig2}
\end{figure}

If we consider Sinister states as a new classification in the hierarchy of quantum correlations, we find they lie between separable and classical states, see Fig.\ref{fig2}. 

Further investigations may fruitfully pursue the relationship between $\mathcal{S}$ and other quantum correlations such as the Negativity \cite{negativity} or the Discord \cite{OllivierZurek02}.  Also the concept might readily be extended to higher dimensional systems.  

\section*{Acknowledgements}
The author would like to thank Alessandro Fedrizzi (whose question inspired this line of investigation), Joe Altepeter, Bernie Englert, Asma Al-Qasimi, Nicolas Quesada, Andrew White and Raymond Fan for useful conversations, and to the Centre for Engineered Quantum Systems at the University of Queensland for hospitality during a visit at which part of this work was conducted.  This work was funded by the Natural Sciences and Engineering Research Council of Canada (RGPIN-2017-06264).

The author would also like to make a special note of regard for the late Emil Wolf, his mentor, collaborator and friend for more than 30 years, and whose 100th birthday this special issue is marking.

\section*{Appendix A: Sinisterness of Separable States}
\renewcommand{\theequation}{{\rm A.\arabic{equation}}}
\setcounter{equation}{0}
\setcounter{subsection}{0}

Using eq.(\ref{detdef}), we find
\begin{equation}
\mbox{Det}\{c_{ij}\}= 
\frac{1}{6}\sum_{k,\ell,m=1}^4  
p_k p_\ell p_m
T_{k \ell m}({\bf a})
T_{k \ell m}({\bf b})
\label{bigsum}
\end{equation}
where
\begin{equation}
T_{k \ell m}({\bf a})= \left[
({\bf a}_{k} -\bar{\bf a})
\times
({\bf a}_{\ell} -\bar{\bf a})\right]\cdot ({\bf a}_{m} -\bar{\bf a})
\label{Tdef}
\end{equation}
represents the scalar triple product. The sum in eq.(\ref{bigsum}) has 64 terms, but it can be readily simplified using the properties of the scalar triple product: firstly, if any of the three indices $k$, $\ell$ or $m$ have the same value, the product is necessarily zero, reducing the sum to 24 terms; these 24 terms can be readily obtained by permuting just four possibilities, viz. $\{k \ell m\}=\{1 2 3\}$, $\{124\}$, $\{134\}$ or $\{234\}$.  The scalar triple product $T_{k \ell m}({\bf a})$ is either invariant or changes sign under permutation of indices; thus the product $T_{k\ell m}({\bf a})T_{k \ell m}({\bf b})$ is invariant under permutation. Since there are 6 permutations of each combination of three indices, the factor of $1/6$ cancels and the sum in eq.(\ref{bigsum}) reduces to just four terms:
\begin{eqnarray}
\mbox{Det}\{c_{i,j}\}=&&
p_1p_2p_3T_{123}({\bf a})T_{123}({\bf b})\nonumber\\
&&+p_1p_2p_4T_{124}({\bf a})T_{124}({\bf b})\nonumber\\
&&+p_1p_3p_4T_{134}({\bf a})T_{134}({\bf b})\nonumber\\
&&+p_2p_3p_4T_{234}({\bf a})T_{234}({\bf b}).
\label{littlesum}
\end{eqnarray}
Expanding the triple product term eq.(\ref{Tdef})
\begin{eqnarray}
T_{k \ell m}({\bf a})=&&
({\bf a}_{k}\times{\bf a}_{\ell})\cdot{\bf a}_{m} 
-(\bar{\bf a}\times{\bf a}_{\ell})\cdot{\bf a}_{m} \nonumber\\
&&-({\bf a}_{k}\times\bar{\bf a})\cdot{\bf a}_{m} 
-({\bf a}_{k}\times{\bf a}_{\ell})\cdot\bar{\bf a}\nonumber\\
= 
\sum_{n=1}^4 p_n&&\left[
({\bf a}_{k}\times{\bf a}_{\ell})\cdot{\bf a}_{m} 
-({\bf a}_{n}\times{\bf a}_{\ell})\cdot{\bf a}_{m}\right. \nonumber\\
&&\left.-({\bf a}_{k}\times{\bf a}_{n})\cdot{\bf a}_{m} 
-({\bf a}_{k}\times{\bf a}_{\ell})\cdot{\bf a}_{n}\right]\nonumber\\
= 
\sum_{n=1}^4 p_n&&
\left[
({\bf a}_{k}-{\bf a}_{\ell})
\times
({\bf a}_{\ell}-{\bf a}_{m})
\right]\cdot 
({\bf a}_{m}-{\bf a}_{n}).\nonumber\\
&&
\label{Tderived}
\end{eqnarray}
It is convenient to define the following scalar product of four vectors:
\begin{equation}
\mathcal{V}({\bf a}_1, {\bf a}_2, {\bf a}_3, {\bf a}_4) = 
\frac{1}{6}
\left[
({\bf a}_1-{\bf a}_2)
\times
({\bf a}_2-{\bf a}_3)
\right]\cdot 
({\bf a}_3-{\bf a}_4).
\label{Vdef}
\end{equation}
This quantity has a simple geometric interpretation: $\left|\mathcal{V}({\bf a}_1, {\bf a}_2, {\bf a}_3, {\bf a}_4)\right|$ is the volume of the irregular tetrahedron whose vertices have the position vectors ${\bf a}_1$, ${\bf a}_2$, ${\bf a}_3$ and ${\bf a}_4$.
One can show by direct evaluation that for $\{k,\ell,m,n\}\in\{1,2,3,4\}$
\begin{equation}
\mathcal{V}({\bf a}_k, {\bf a}_\ell, {\bf a}_m, {\bf a}_n) = 
s_{k \ell m n} \mathcal{V}({\bf a}_1, {\bf a}_2, {\bf a}_3, {\bf a}_4),
\end{equation}
where $s_{k \ell m n}$ is the signature of the permutation $\{k,\ell,m,n\}$ with respect to $\{1,2,3,4\}$ (i.e. $s_{k \ell m n}=1$ if $\{k,\ell,m,n\}$ is an even permutation of $\{1,2,3,4\}$ and $s_{k \ell m n}=-1$ if $\{k,\ell,m,n\}$ is an odd permutation of $\{1,2,3,4\}$). This immediately implies that $\mathcal{V}({\bf a}_k, {\bf b}_\ell, {\bf c}_m, {\bf d}_n) = 0$ if any two indices are the same.  Substituting from eq.(\ref{Vdef}) into eq.(\ref{Tderived}) we find
\begin{eqnarray}
T_{k \ell m}({\bf a})&=& 
6 \sum_{n=1}^4p_n \mathcal{V}({\bf a}_k, {\bf a}_\ell, {\bf a}_m, {\bf a}_n),\nonumber\\
&=& 
6 \left(\sum_{n=1}^4 s_{k \ell m n} p_n\right) \mathcal{V}({\bf a}_1, {\bf a}_2, {\bf a}_3, {\bf a}_4).
\end{eqnarray}
Substituting into eq.(\ref{littlesum}), we find:
\begin{eqnarray}
\mbox{Det}\{c_{ij}\}&=&
36\left[(s_{1234})^2 p_1p_2p_3p_4^2+
(s_{1243})^2 p_1p_2p_4p_3^2\right.\nonumber\\
&&\left.+(s_{1342})^2 p_1p_3p_4p_2^2+(s_{2341})^2 p_2p_3p_4p_1^2\right]\nonumber\\
&&\times 
\mathcal{V}({\bf a}_1, {\bf a}_2, {\bf a}_3, {\bf a}_4)
\mathcal{V}({\bf b}_1, {\bf b}_2, {\bf b}_3, {\bf b}_4)\nonumber\\
&=&36p_1p_2p_3p_4\nonumber\\
&&\times\mathcal{V}({\bf a}_1, {\bf a}_2, {\bf a}_3, {\bf a}_4)
\mathcal{V}({\bf b}_1, {\bf b}_2, {\bf b}_3, {\bf b}_4),
\label{littlesum2}
\end{eqnarray}
where we have used the fact that $(s_{k \ell m n})^2=1$ and $p_1+p_2+p_3+p_4=1$ to derive the final line.  

\section*{Appendix B: Maximum Volume of a Pyramid with Vertices on a Sphere}
\begin{figure}[h]
\centerline{\includegraphics[width=6cm]{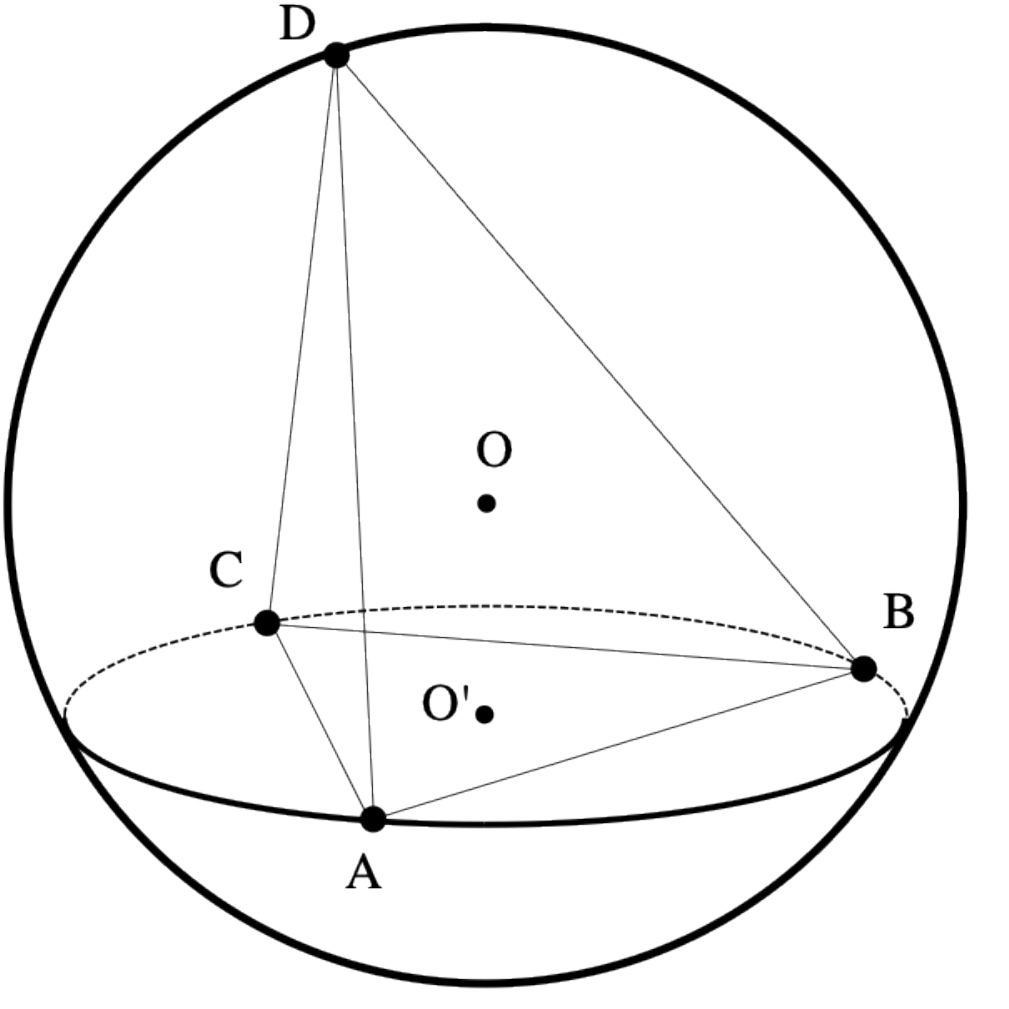}}
\caption{Diagram of the notation used in Appendix B.}
\label{fig3}
\end{figure}

\renewcommand{\theequation}{{\rm B.\arabic{equation}}}
\setcounter{equation}{0}
\setcounter{subsection}{0}
Here we provide a proof for the formula for the maximum volume of the tetrahedron whose vertices $A$, $B$, $C$ and $D$ all lie on the surface of a sphere.  Let $O$ be the center of the sphere (of radius $R$) and $O^\prime$ the center of the circumscribed circle $K_{ABC}$ of the triangle $ABC$ (see Fig,(\ref{fig3})).  

The volume of the pyramid $ABCD$ is $\mathscr{V}_{ABCD}=\frac{1}{3} h_D\mathscr{A}_{ABC}$, where $h_D$ is the perpendicular height of $D$ above the $ABC$ plane, and $\mathscr{A}_{ABC}$ is the area of the triangle $ABC$.  In terms of the vector product, $\mathscr{A}_{ABC}=\frac{1}{2}|\overrightarrow{AB}\times \overrightarrow{BC}|$, while, using the scalar product $h_D= \overrightarrow{CD}\cdot (\overrightarrow{AB}\times \overrightarrow{BC})/|\overrightarrow{AB}\times \overrightarrow{BC}|$; thus $\mathscr{V}_{ABCD}=\frac{1}{6} |\overrightarrow{CD}\cdot (\overrightarrow{AB}\times \overrightarrow{BC})|=\frac{1}{6} |[(\overrightarrow{OA}-\overrightarrow{OB})\times (\overrightarrow{OB}-\overrightarrow{OC})]\cdot(\overrightarrow{OC}-\overrightarrow{OD})|$.

To find the maximum value of $\mathscr{V}_{ABCD}$ first, let us show that $\overrightarrow{O^\prime O}$ is perpendicular to the plane $ABC$. Let the $z$-axis be perpendicular to the plane $ABC$, so that any point $P$ on the circumference of $K_{ABC}$ has the position vector $\overrightarrow{O^\prime P}=r(\cos\phi,\sin\phi,0)$, where $r$ is the radius of $K_{ABC}$.  Using the same coordinate axes, let $\overrightarrow{O^\prime O}=a(\sin\theta_0\cos\phi_0,\sin\theta_0\sin\phi_0,\cos\theta_0)$, and hence $|\overrightarrow{O^\prime P}-\overrightarrow{O^\prime O}|=\sqrt{r^2+a^2- 2 a r \sin\theta_0\cos(\phi-\phi_0)}$.  Since all points on $K_{ABC}$ are on the surface of the sphere it follows that $|\overrightarrow{O^\prime P}-\overrightarrow{O^\prime O}|=R$, for all values of $\phi$; hence differentiating the square root term with respect to $\phi$ we find $\sin\theta_0=0$ and hence $\overrightarrow{O^\prime O}=(0,0,a)$, i.e. parallel to the z-axis and hence perpendicular to the plane $ABC$.

Now we shall demonstrate the maximum value of $h_D$ is $a+R$, which occurs when $\overrightarrow{OD}\parallel\overrightarrow{O^\prime O}$.  If we write $\overrightarrow{OD}=R(\sin\theta_d\cos\phi_d,\sin\theta_d\sin\phi_d,\cos\theta_d)$ and $\overrightarrow{O^\prime O}=(0,0,a)$ we find $h_D=\overrightarrow{O^\prime D}\cdot \vec{e}_z = a+R\cos\theta_d$.  Thus, differentiating to find the maximum, we find $\partial h_D/\partial \theta_d =0$ and $\partial^2 h_D/\partial \theta_d^2 < 0$ when $\theta_d=0$, implying $\overrightarrow{OD}_{max}= R(0,0,1)$ and $h_{D,max}=a+R$.  

\begin{figure}[h]
\centerline{\includegraphics[width=6cm]{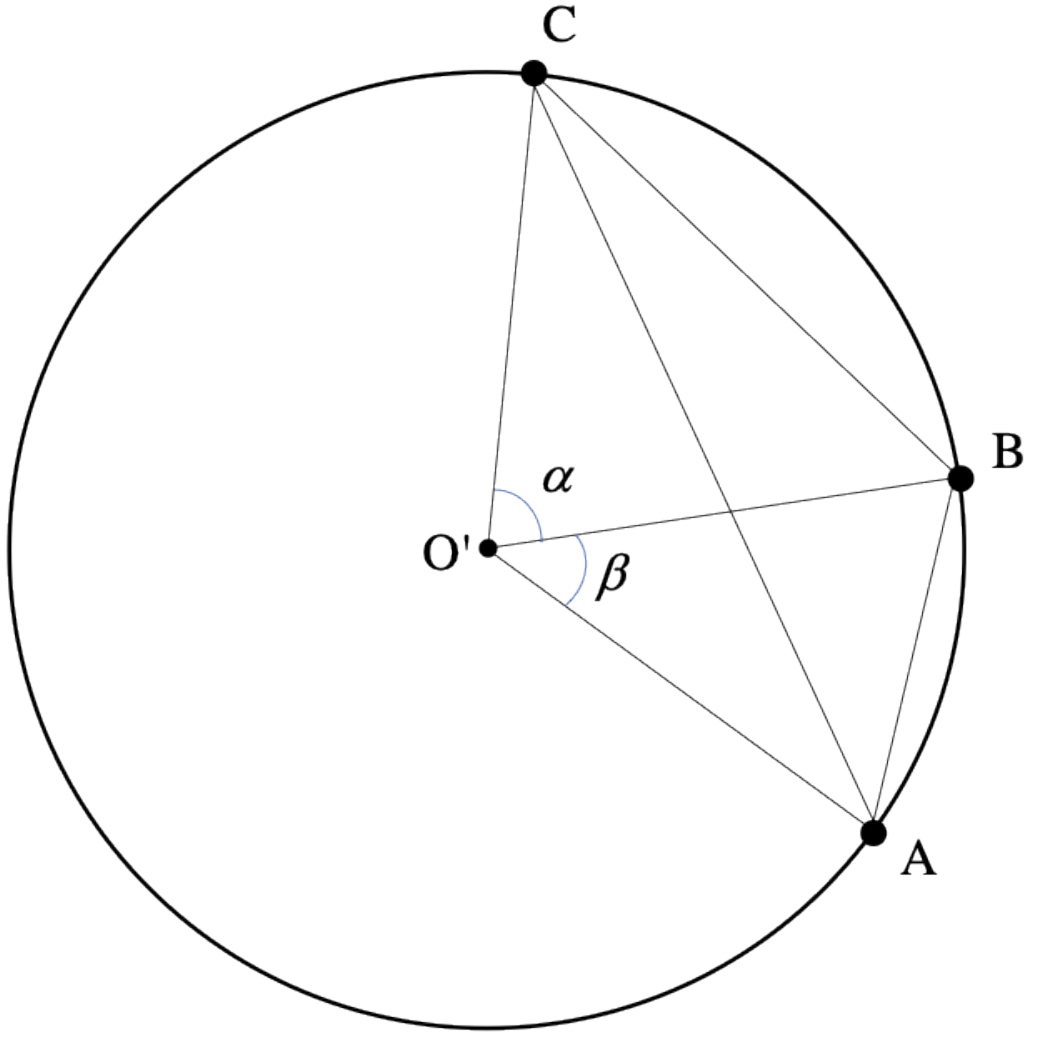}}
\caption{Diagram of the notation used to find $\mathscr{A}_{ABC}$.}
\label{fig4}
\end{figure}

The next step is to maximize the area of the triangle $ABC$.  Intuition tells us that the maximum will occur in the symmetric situation, i.e. $ABC$ is an equilateral triangle.  However intuition, like sincerity, is always subject to proof, which we now supply.  Referring to Fig.(\ref{fig4}), the area $\mathscr{A}_{ABC}=\mathscr{A}_{ABO^\prime}+\mathscr{A}_{BCO^\prime}-\mathscr{A}_{ACO^\prime}$ $= \frac{1}{2} r^2[\sin(\alpha)+\sin(\beta)-\sin(\alpha+\beta)]$, where $r$ is the radius of the circumscribed circle, $\alpha$ is the angle $AO^\prime B$ and $\beta$ is the angle $BO^\prime C$, as shown.  Differentiating $\mathscr{A}_{ABC}$, we find the condition for a maximum:
\begin{eqnarray}
\frac{\partial}{\partial \alpha}\mathscr{A}_{ABC} &=&\frac{r^2}{2}[\cos(\alpha)-\cos(\alpha+\beta)]=0
\label{parAa}\\
\frac{\partial}{\partial \beta}\mathscr{A}_{ABC} &=&\frac{r^2}{2}[\cos(\beta)-\cos(\alpha+\beta)]=0.
\label{parAb}
\end{eqnarray}
These equations imply that, at a maximum or minimum, $\cos(\alpha)=\cos(\beta)$, and hence $\beta= \pm\alpha$, (where we assume $\alpha$ and $\beta$ have values between $0$ and $2 \pi$ .  If $\beta+\alpha= 0$, eqs.(\ref{parAa}-\ref{parAb}) imply that $\cos(\alpha)=\cos(\beta)=1$, and $\sin(\alpha)=\sin(\beta)=0$ and hence the area $\mathscr{A}_{ABC}=0$.  Similarly, if $\beta= \alpha$, eq.(\ref{parAa}) implies $\cos(\alpha)=\cos(2 \alpha)$, and hence $\alpha = 2 \alpha  $, i.e. $\alpha = 0 $ (which again, implies $\mathscr{A}_{ABC}=0$), or $\alpha= 2\pi- 2\alpha$, or $\alpha = \beta= 2 \pi/3 $.  This last case corresponds to the equilateral triangle; it can be confirmed to be the maximum by showing that the eigenvalues of the Hessian are all negative. Since $\sin(2\pi/3)=\sqrt{3}/2$ and $\sin(4\pi/3)=-\sqrt{3}/2$, we find $\mathscr{A}_{ABC, max}=3\sqrt{3}r^2/4$.

The last step is to maximize the volume $\mathscr{V}_{ABCD,max} = \frac{1}{3} \mathscr{A}_{ABC, max} h_{D,max}=a+R$.  The only remaining underdetermined parameter is $a$, the length $OO^\prime$.  By Pythagoras's Theorem, $r^2=R^2-a^2$, and hence $\mathscr{V}_{ABCD,max}(a)=(a+R)(R^2-a^2)\sqrt{3}/4$.  Differentiating, we find $\frac{\partial}{\partial a} \mathscr{V}_{ABCD,max}=0$ when $a=-R$ or $a=R/3$.  The former solution corresponds to a minimum (volume zero), while the second corresponds to the maximum.  Hence we finally obtain $\mathscr{V}_{ABCD, max}= R^3 8/9\sqrt{3}$.  Using this solution, it can be shown reasonably straightforwardly that the distances between any two of the points $A$, $B$, $C$ and $D$ is the same, viz., $R \sqrt{8/3}$, and thus the maximum volume corresponds to a regular tetrahedron.

\section*{Appendix C: Useful properties of Non-Hermitian Matrices}
\renewcommand{\theequation}{{\rm C.\arabic{equation}}}
\setcounter{equation}{0}
\setcounter{subsection}{0}
In this appendix we give a brief review some of the key properties of Non-Hermtian matrices and the associated perturbation series required to calculate variation in the eigenvalues.  A more thorough exposition can be found in, for example, refs. \cite{MorseFesh, Brody}.

\subsection{Eigenvectors and Bi-othogonality}
Let ${\rm M}$ be a non-Hermtian matrix. It will have both left and right eigenvectors, viz.,
\begin{eqnarray}
{\rm M}\ket{\rm v}&=&\mu\ket{\rm v}\label{reveqn},\\
\bra{\rm u}{\rm M}&=&\mu\bra{\rm u}\label{leveqn}.
\end{eqnarray}
One might initially suspect that the left and right eigenvectors have different associated eigenvalues; however both eq.(\ref{reveqn}) and eq.(\ref{leveqn}) imply the same characteristic equation, viz., $\mbox{Det}\{{\rm M}-r\}=0$, and thus the left and right eigenvalues are identical.  Since ${\rm M}$ is a 4x4 matrix, the characteristic equation is a quartic polynomial equation and there are at most 4 eigenvalues, $r_n$, with corresponding eigenvectors $\bra{\rm u_n}$ and $\ket{\rm v_n}$; it is useful to normalize these vectors by $\braket{\rm u_n}{\rm v_n}=1$, and, in general one may {\em not} assume $\braket{\rm u_n}{\rm u_n}=1$ or $\braket{\rm v_n}{\rm v_n}=1$; we denote vectors normalized in this manner by lower-case Latin symbols with a Roman typeface (e.g. ${\rm u}, {\rm v}$, etc), as opposed to Greek symbols used for vectors with the standard normalization (i.e. $\braket{\varphi}{\varphi}=1$).  

Equations (\ref{reveqn}) and  (\ref{leveqn}) imply $\bra{\rm u_n}{\rm M}\ket{\rm v_m}=\mu_m\braket{\rm u_n}{\rm v_m}=\mu_n\braket{\rm u_n}{\rm v_m}$, and thus
\begin{equation}
(\mu_m-\mu_n)\braket{\rm u_n}{\rm v_m}=0
\end{equation}
For non-degenerate systems, the appropriately normalized left and right eigenvectors are {\em bi-orthogonal}, i.e. 
\begin{equation}
\braket{\rm u_n}{\rm v_m}=\delta_{mn}. \label{biorth}
\end{equation}
Note however neither the left nor the right sets of eigenvectors are in general  orthogonal (i.e. neither $\braket{\rm u_m}{\rm u_n}$ nor $\braket{\rm v_m}{\rm v_n}$ need be zero if $m\neq n$).  Provided there exists a state $\ket{\rm u_n}$ such that $\braket{\rm u_n}{\rm v_n}\neq0$ (this cannot be proved in general, see \cite{SternWalk}, Section III), both sets of four eigenvectors span the 4 dimensional space, and thus we find:
\begin{eqnarray}
\sum_{m=1}^4\ket{\rm v_m}\bra{\rm u_m} &=&{\rm I} \label{rlident},\\
\sum_{m=1}^4\ket{\rm u_m}\bra{\rm v_m} &=&{\rm I} \label{lrident}.
\end{eqnarray}

\subsection{Bi-othogonal Perturbation Theory}
The perturbation theory for non-Hermitian matrices has been discussed by various authors (e.g. \cite{SternWalk,JMKW00, Brody}); here we give a brief summary of the important results. Suppose we wish to find the eigenvalues and eigenvectors of a matrix ${\rm M}+\lambda\,{\delta\rm M}$ to the first power of the parameter $\lambda$.  As usual, one assumes that the eigenvalues and eigenvectors can be expended in a power series, i.e.
\begin{widetext}
\begin{equation}
\left\{{\rm M}-\mu_n+\lambda (\delta{\rm M} -\delta\mu_n)+O[\lambda^2]\right\}
\left\{\ket{\rm v_n}+\lambda\ket{\delta{\rm v}_n}+O[\lambda^2]\right\}=0.
\end{equation}
\end{widetext}
This equality must be true for arbitrary values of $\lambda$; hence
expanding this expression and collecting terms in $\lambda$ we obtain:
\begin{eqnarray}
&&\left({\rm M} -\mu_n\right)\ket{\rm v_n}=0 \label{pertM0}\\
&&\left({\rm M} -\mu_n\right)\ket{\delta{\rm v}_n}+\left(\delta{\rm M} -\delta\mu_n\right)\ket{\rm v^{(0)}_n}=0\label{pertM1}
\end{eqnarray}
and so on.  
Equation (\ref{pertM0}) implies that $\ket{{\rm v}_n}$, $(n=1,2,3,4)$ are the right eigenvectors of ${\rm M}$; together with the corresponding left eigenvectors $\bra{{\rm u}_n}$, they form a bi-orthogonal system.  Applying $\bra{{\rm u}_m}$ to the left of eq.(\ref{pertM1}) we find, after some re-arrangement
\begin{equation}
\delta\mu_n=\bra{{\rm u}_m}\delta{\rm M}\ket{{\rm v}_m},
\label{deltamu}
\end{equation}
and 
\begin{equation}
\ket{\delta{\rm v}_n}=\braket{{\rm u}_n}{\delta{\rm v}_n}\ket{{\rm v}_n}+
\sum_{\stackrel{\scriptstyle m}{m \neq n}}
\frac{\bra{{\rm u}_m}\delta{\rm M}\ket{{\rm v}_n}}
{\mu_n-\mu_m}\ket{{\rm v}_m}.\label{ketdeltavM}
\end{equation}
Normalization requires that $\braket{{\rm u}_n}{\delta{\rm v}_n}=-\braket{\delta{\rm u}_n}{{\rm v}_n}$, but otherwise the form of this coefficient is undetermined in general.  Higher order terms can be obtained by continuing the expansion to higher powers of $\lambda$, but these will not be needed for the current analysis.

\section*{Appendix D: Quantum Concurrence}
\renewcommand{\theequation}{{\rm D.\arabic{equation}}}
\setcounter{equation}{0}
\setcounter{subsection}{0}

Here we briefly review some relevant properties of the concurrence of mixed two-qubit quantum states.  For more details the reader is referred to the seminal work of Wootters \cite{Wootters98}.

\subsection{Pure States}
For a pure two qubit state $\ket{\psi}=\alpha\ket{00}+\beta\ket{01}+\gamma\ket{10}+\delta\ket{11}$, the entanglement is well characterized by the pure state Concurrence $\mathcal{C}(\psi)=2|\alpha\delta-\beta\gamma|$.  When $\mathcal{C}(\psi)=0$, the state may be written as a product for two single qubit states; when $\mathcal{C}(\psi)=1$, the state is maximally entangled.

The ``spin-flipped" state is defined as
\begin{equation}
\ket{\tilde{\psi}}=(\sigma_2\otimes\sigma_2)\ket{\psi^*},
\end{equation}
where $\ket{\psi^*}$ is related to the state $\ket{\psi}$ by complex-conjugation of the probability amplitudes {\em in the computational basis}; the basis states themselves are implicitly assumed to be real-valued and are unaffected.  Thus 
\begin{eqnarray}
\ket{\tilde{\psi}}&=&\delta^*\ket{00}-\gamma^*\ket{01}-\beta^*\ket{10}+\alpha^*\ket{11},\\
\bra{\tilde{\psi}}&=&\delta\bra{00}-\gamma\bra{01}-\beta\bra{10}+\alpha\bra{11}.
\end{eqnarray}
Note that for any two pure states $\ket{\psi}$ and $\ket{\phi}$  $\braket{\tilde{\psi}}{\phi}=\braket{\tilde{\phi}}{\psi}$; also $|\braket{\tilde{\psi}}{\psi}|=\mathcal{C}(\psi)$ is the pure state Concurrence.

\subsection{Maximally Entangled States}
Any maximally entangled state $\ket{\psi_{ME}}$, by definition, has the property $\mathcal{C}(\psi_{ME})=1$; since $\ket{\psi_{ME}}$ is normalized in the usual manner (i.e. $\braket{\psi_{ME}}{\psi_{ME}}=1$), a maximally entangled state must have the property $\ket{\tilde{\psi_{ME}}}=\exp(i \phi)\ket{\psi_{ME}}$, where $\phi$ is some phase.  Since, for any state, $\ket{\accentset{\approx}{\psi}}=\ket{\psi}$, we see that $\exp(i \phi)=\pm1$, and thus the projector $\Pi_{ME}=\ket{\psi_{ME}}\bra{\psi_{ME}}$ of a maximally entangled state has the property 
\begin{equation}
\tilde{\Pi}_{ME}=\Pi_{ME}.
\end{equation}

\subsection{Mixed States}
A mixed state is conventionally represented by a density operator $\rho$ , which can be written as the sum of projectors of pure states weighted by some probability, viz.,
\begin{equation}
\rho=\sum_n p_n\ket{\psi_n}\bra{\psi_n},
\end{equation}
where $\sum_n p_n=1$ and $0<p_n\le1$. The number of terms in the sum must be greater or equal to the rank of $\rho$. Since there is no assumption of orthogonality of $\ket{\psi_n}$, this decomposition (termed a {\em convex hull}) is not the same as an eigenstate decomposition; indeed an infinite number of such pure state ensembles exist for any given density operator.

The obvious approach to finding the mixed-state Concurrence would be to average the pure-state concurrences of the states making up the convex hull, i.e. $\overline{\mathcal{C}}=\sum_n p_n {C}(\psi_n)$.  However, because the convex hull decomposition is not unique, this average can take a range of values; for example, any density operator which is diagonal in the Bell state basis can be decomposed by choosing a maximally entangled basis in which  $\overline{\mathcal{C}}=1$; in particular, one can obtain $\overline{\mathcal{C}}=1$ for the maximally mixed state.  Thus to faithfully characterize the entanglement of a mixed state one chooses the mixed state decomposition for which $\overline{\mathcal{C}}$ is a minimum, viz:
\begin{equation}
{\mathcal{C}}(\rho)=\min_{\left\{\psi_n\right\}}\sum_n p_n{C}(\psi_n).
\end{equation}
This minimum can be shown to be equal to the following expression
\begin{equation}
{\mathcal{C}}(\rho)=\mbox{max}\{\lambda_1-\lambda_2-\lambda_3-\lambda_4, 0\},
\label{ConcDef1}
\end{equation}
where $\lambda_n$ ($n=1,\ldots 4$, numbered in decreasing order, so $\lambda_1$ is the largest, $\lambda_4$ is the smallest) are the the eigenvalues of the matrix $\mathcal{R}= \sqrt{\sqrt{\rho}\tilde{\rho}\sqrt{\rho}}$, $\tilde{\rho}$ being the ``spin-flipped'' density matrix, $\tilde{\rho}= (\sigma_2\otimes\sigma_2) \rho^* (\sigma_2\otimes\sigma_2)$. As with the pure state, the complex-conjugate of the density operator $\rho^*$ is defined in the standard computational basis (this operation is basis dependent: if, for example, we did a complex conjugation in the eigenbasis of $\rho$, it would not have any effect).

\subsection{Properties of the Bi-orthogonal Eigenvectors}
Alternatively (and in the present problem, fortuitously, since the square roots of non-commuting operators in the expression for $\mathcal{R}$ are awkward to manipulate analytically) one can 
show that eq.(\ref{ConcDef1}) is equivalent to 
\begin{equation}
{\mathcal{C}}(\rho)=\mbox{max}\{\sqrt{r_1}-\sqrt{r_2}-\sqrt{r_3}-\sqrt{r_4}, 0\},
\label{ConcDef2}
\end{equation}
where $r_n$ ($n=1,...4$, again numbered in decreasing magnitude) are the eigenvectors of the {\em non-Hermitian} matrix defined by the formula
\begin{equation}
{\mathscr R}= \tilde{\rho}\rho.
\label{Rdef}
\end{equation}
Relevant properties of non-Hermitian matrices and their bi-orthogonal expansions are reviewed briefly in Appendix C.
Let us denote the left and right eigenvectors of ${\mathscr R}$ by $\bra{\rm u_n}$ and $\ket{\rm v_n}$ respectively; they are normalized in the way described in Appendix C, i.e. $\braket{\rm u_n}{\rm v_n}=\delta_{mn}$.  Thus
\begin{eqnarray}
\bra{\rm u_n}{\mathscr R}&=&r_n\bra{\rm u_n}\label{levR1} \\
{\mathscr R}\ket{\rm v_n}&=&r_n\ket{\rm v_n} \label{revR1}
\end{eqnarray}
The transpose of eq.(\ref{revR1}) gives
\begin{equation}
\bra{\rm v_n^*}{\mathscr R}^T=r_n\bra{\rm v_n^*},
\end{equation}
where ${\mathscr R}^T=\rho^T (\sigma_2\otimes\sigma_2)\rho (\sigma_2\otimes\sigma_2) =(\sigma_2\otimes\sigma_2){\mathscr R}(\sigma_2\otimes\sigma_2)$.  Hence
\begin{equation}
\bra{\rm v_n^*}(\sigma_2\otimes\sigma_2){\mathscr R}=r_n\bra{\rm v_n^*}(\sigma_2\otimes\sigma_2).
\end{equation}
Thus the left eigenvectors of ${\mathscr R}$ are just the ``spin-flipped" right eigenvectors $\ket{\rm u_n}=\ket{\tilde{\rm v}_n}$, and we obtain
\begin{equation}
\braket{\tilde{\rm v}_m}{{\rm v}_n}=\delta_{mn}
\label{etatildexi}.
\end{equation}
\noindent
If we now took the Hermitian adjoint of eq.(\ref{revR1}), and use the fact that both $\rho$ and $\tilde{\rho}$ are both Hermitian, we obtain:
\begin{equation}
\bra{\rm v_n}\rho\tilde{\rho}=r_n^*\bra{\rm v_n}.
\end{equation}
Multiplying both sides on the right by $\rho$, we find
\begin{equation}
\bra{\rm v_n}\rho {\mathscr R}=r_n^*\bra{\rm v_n}\rho.
\end{equation}
Comparing with eq.(\ref{levR1}) we see this implies that the eigenvalues $r_n$ are all real, and that, after normalizing,
\begin{equation}
\ket{\tilde{\rm v}_n} =\frac{\rho\ket{\rm v_n}}{\bra{\rm v_n}\rho\ket{\rm v_n}}.
\end{equation}
The bi-orthonality property eq.(\ref{biorth}) thus implies 
\begin{equation}
\bra{\rm v_n}\rho\ket{\rm v_m}=\delta_{mn}\bra{\rm v_n}\rho\ket{\rm v_n} \label{rhoxidiag}.  
\end{equation}
Although in a sense we have diagonalized $\rho$, this equation does {\em not} imply $\ket{\rm v_n}$ is an eigenvector of $\rho$, since the vectors $\ket{\rm v_n}$ are not orthogonal, as would be the case for the eigenvectors of a Hermitian operator.  The eigenvalues $r_n$ are, from eqs.(\ref{reveqn}),(\ref{lrident}) and (\ref{etatildexi}), given by 
\begin{eqnarray}
r_n&=&\bra{\tilde{\rm v}_n}{\mathscr R}\ket{\rm v_n}\nonumber\\
&=&\sum_m
\bra{\tilde{\rm v}_n}\tilde{\rho}\ket{\tilde{\rm v}_m}\bra{\rm v_m}\rho\ket{\rm v_n}\nonumber\\
&=&\bra{\tilde{\rm v}_n}\tilde{\rho}\ket{\tilde{\rm v}_n}\bra{\rm v_n}\rho\ket{\rm v_n}.
\end{eqnarray}
Since $(\sigma_2\otimes\sigma_2)^2={\rm I}$, matrix element $\bra{\tilde{\rm v}_n}\tilde{\rho}\ket{\tilde{\rm v}_n}$ is equal to $\bra{\rm v_n^*}\rho^*\ket{\rm v^*_n}$, which, because of the Hermiticity of $\rho$, reduces to $\bra{\rm v_n}\rho\ket{\rm v_n}$.  Hence the eigenvalues $r_n$ are all real non-negative numbers, and
\begin{equation}
\bra{\rm v_m}\rho\ket{\rm v_n} = \sqrt{r_n}\delta_{mn}.
\label{rootr1}
\end{equation} 
\vspace{5mm}

\subsection{Perturbation Analysis}
If the density operator is perturbed, i.e. $\rho\rightarrow \rho+\lambda \delta\rho$, the Concurrence will also be perturbed, ${\mathcal{C}}(\rho+\lambda \delta\rho)= {\mathcal{C}}(\rho)+\lambda \delta\mathcal{C} + O[\lambda^2]$.  Here we will derive an expression for the first order change in Concurrence, $\delta\mathcal{C}$.  Note that $\delta\rho$ is constrained to be Hermitian and trace zero, and that $\rho+\lambda \delta\rho$ is non-negative definite.  We also implicitly assume that ${\mathcal{C}}(\rho)\neq0$, so this analysis should be treated with caution for states that are close to being separable. The first order perturbation to the matrix ${\mathscr R}$ is $\delta{\mathscr R}=\tilde{\rho}(\delta\rho)+(\delta\tilde{\rho})\rho$, and the first order perturbation to the eigenvectors is given by eq.(\ref{ketdeltavM}):
\begin{equation}
\ket{\delta{\rm v}_n}=\braket{\tilde{\rm v}_n}{\delta{\rm v}_n}\ket{{\rm v}_n}+
\sum_{\stackrel{\scriptstyle m}{m \neq n}}
\frac{\bra{\tilde{\rm v}_m}\delta{\mathscr R}\ket{{\rm v}_n}}
{r_n-r_m}\ket{{\rm v}_m}.
\label{dketv1}
\end{equation}
Normalization of the perturbed eigenvectors implies that $\braket{\tilde{\rm v}_n}{\delta{\rm v}_n}+\braket{\delta\tilde{\rm v}_n}{{\rm v}_n}=0$; using $\braket{\tilde{\phi}}{\psi}=\braket{\tilde{\psi}}{\phi}$, we see that this implies $2\braket{\delta\tilde{\rm v}_n}{{\rm v}_n}=0$, hence the first term on the right hand side of eq.(\ref{dketv1}) must be zero. Using eq.(\ref{rootr1}), we find
\begin{widetext}
\begin{eqnarray}
\sqrt{r_n}+\lambda\delta(\sqrt{r_n})&=& 
\left\{\bra{\rm v_n}+\lambda \bra{\delta\rm v_n}\right\}\left\{\rho+\lambda\delta\rho\right\}\left\{\ket{\rm v_n}+\lambda \ket{\delta\rm v_n}\right\}+O[\lambda^2]\nonumber\\
&=&
\bra{\rm v_n}\rho\ket{\rm v_n}+
\lambda\left\{
\bra{\rm v_n}\delta\rho\ket{\rm v_n}+
\bra{\delta\rm v_n}\rho\ket{\rm v_n}+
\bra{\rm v_n}\rho\ket{\delta\rm v_n}\right\}+O[\lambda^2]
\label{rootr2}
\end{eqnarray} 
\end{widetext}
From eq.(\ref{dketv1}) we find
\begin{eqnarray}
\bra{{\rm v}_n}\rho\ket{\delta{\rm v}_n}&=&
\sum_{\stackrel{\scriptstyle m}{m \neq n}}
\frac{\bra{\tilde{\rm v}_m}\delta{\mathscr R}\ket{{\rm v}_n}}
{r_n-r_m}
\bra{{\rm v}_n}\rho\ket{{\rm v}_m}\nonumber\\
&=&0,
\label{dketv2}
\end{eqnarray}
where we have used eq.(\ref{rootr1}); a similar analysis shows that $\bra{\delta{\rm v}_n}\rho\ket{{\rm v}_n}=0$.  Hence we obtain $\delta(\sqrt{r_n})=\bra{\rm v_n}\delta\rho\ket{\rm v_n}$, which implies the following expression for the variation in the Concurrence:
\begin{equation}
\delta\mathcal{C}=\mbox{Tr}\{{\rm W}\delta\rho\}.
\label{deltaC}
\end{equation}
where
\begin{equation}
{\rm W}=
\ket{\rm v_1}\bra{\rm v_1}-
\ket{\rm v_2}\bra{\rm v_2}-
\ket{\rm v_3}\bra{\rm v_3}-
\ket{\rm v_4}\bra{\rm v_4}.
\label{Wdef}
\end{equation}
To re-iterate: $\ket{\rm v_n}$ is the right-eigenvector of the operator ${\mathscr R}=\tilde{\rho}\rho$, normalized so that $\bra{\rm v_n^*}(\sigma_2\otimes\sigma_2)\ket{\rm v_m}=\delta_{mn}$; these vectors do not form an orthonormal basis, and in general the terms in the definition of eq.(\ref{Wdef}) are {\em not} projectors.

For the specific case of the Werner state, we have the simplification that $\mathscr{R}_W$ is Hermitian, eq.(\ref{RWdef}), and thus the bi-orthogonal basis states are also orthogonal.  In this case eq.(\ref{Wdef}) reduces to ${\rm W}=2 \Pi-\rm{I}$, and hence writing $\delta\rho=\rho^\prime-\rho_W$ we find
\begin{equation}
\delta\mathcal{C}_W=2 \left(\mbox{Tr}\{\Pi\rho^\prime\}-\frac{(3\epsilon+1)}{4}\right),
\label{deltaCW}
\end{equation}
where we have used $\mbox{Tr}\{\Pi\rho_W\}=(3\epsilon+1)/4$.

\section*{Appendix E: Expansion of Determinants and Perturbation Theory for the Sinisterness}
\renewcommand{\theequation}{{\rm E.\arabic{equation}}}
\setcounter{equation}{0}
\setcounter{subsection}{0}
Here we prove a useful expansion formula for determinant of the form
\begin{equation}
\mathcal{D}(\lambda)=\mbox{Det}\{{\rm I}+\lambda{\rm A}\},
\label{D1}
\end{equation}
where $\lambda$ is an expansion parameter.  Introducing the eigenvalues of ${\rm A}$, $\{a_n\}$, eq.(\ref{D1}) can be re-written as follows:
\begin{equation}
\mathcal{D}(\lambda)=\prod_{n=1}^N (1+\lambda a_n).
\label{D2}
\end{equation}
Taking the logarithm, we find
\begin{eqnarray}
\ln(\mathcal{D}(\lambda))&=&\sum_{n=1}^N \ln(1+\lambda a_n)\nonumber\\
&=&\sum_{n=1}^N \sum_{m=1}^\infty -\frac{(-\lambda a_n)^m}{m}\nonumber\\
&=&\sum_{m=1}^\infty -\frac{(-\lambda )^m}{m}\mbox{Tr}\{{\rm A}^m\}.
\label{D3}
\end{eqnarray}
Taking the exponential of both sides we find
\begin{widetext}
\begin{eqnarray}
\mathcal{D}(\lambda)&=&\exp\left(\lambda \mbox{Tr}\{{\rm A}\} \right)
\exp\left(-\frac{\lambda^2}{2} \mbox{Tr}\{{\rm A}^2\} \right)
\exp\left(\frac{\lambda^3}{3} \mbox{Tr}\{{\rm A}^3\} \right)
\exp\left(-\frac{\lambda^4}{4} \mbox{Tr}\{{\rm A}^4\} \right)+O[\lambda^5]\nonumber\\
&=&1+\lambda \mbox{Tr}\{{\rm A}\} +
\frac{\lambda^2}{2} \left(\mbox{Tr}\{{\rm A}\}^2-\mbox{Tr}\{{\rm A^2}\} \right)
+
\frac{\lambda^3}{6} \left(\mbox{Tr}\{{\rm A}\}^3-3\,\mbox{Tr}\{{\rm A}\}\mbox{Tr}\{{\rm A^2}\}+2\,\mbox{Tr}\{{\rm A^3}\} \right)\nonumber\\
&&+
\frac{\lambda^4}{24} \left(\mbox{Tr}\{{\rm A}\}^4-6\,\mbox{Tr}\{{\rm A}\}^2\mbox{Tr}\{{\rm A^2}\}+
3\,\mbox{Tr}\{{\rm A^2}\}^2+8\,\mbox{Tr}\{{\rm A}\}\mbox{Tr}\{{\rm A^3}\}-
6\,\mbox{Tr}\{{\rm A^4}\} \right)
+O[\lambda^5].
\label{D5}
\end{eqnarray}
\end{widetext}
For 4x4 matrices, the series terminates at the fifth ($\propto \lambda^4$) term; the coefficients of $\lambda^5$ and higher powers are zero for $N=4$.

This series can be used to calculate the variation of the Sinisterness $\mathcal{S}$.  From eq.(\ref{SinisDef}) we have
\begin{equation}
\mathcal{S}(\rho)=-16\mbox{Det}\{\mathscr{G}(\rho)\}
\end{equation}
where $\mathscr{G}(\rho)$ is the partial transposition of the density matrix $\rho $ given by eq.(\ref{Gdef}).  If we now consider the variation of the density matrix $\rho\rightarrow \rho + \lambda \delta\rho$, we find:
\begin{eqnarray}
\mathcal{S}(\rho + \lambda \delta\rho) &=& -16\mbox{Det}\{\mathscr{G}(\rho)+\lambda\mathscr{G}(\delta\rho) \}\nonumber\\
&=& -16\mbox{Det}\{\mathscr{G}(\rho)\}\mbox{Det}\{{\rm I}+\lambda\mathscr{G}(\rho)^{-1}\mathscr{G}(\delta\rho) \}\nonumber\\
&=& \mathscr{G}(\rho) \left(1+\lambda\mbox{Tr}\{\mathscr{G}(\rho)^{-1}\mathscr{G}(\delta\rho)\} +\mbox{O}[\lambda^2]\right).\nonumber\\
\end{eqnarray}
For the specific case of the Werner state $\rho_W=(\frac{1-\epsilon}{4})\rm{I}+\epsilon \Pi$, we find $\mathscr{G}_W=\frac{\epsilon}{2}\rm{I}+(\frac{1-\epsilon}{2})\Pi$ and $\mathscr{G}^{-1}_W=\frac{2}{\epsilon}[\rm{I}+(\epsilon-1)\Pi]$.  Writing $\delta\rho=\rho^\prime-\rho$, we find
\begin{eqnarray}
\mbox{Tr}\{\mathscr{G}^{-1}_W\mathscr{G}(\delta\rho)\} &=& \mbox{Tr}\{\mathscr{G}^{-1}_W\mathscr{G}(\rho^\prime)-\rm{I}\}\nonumber\\
&=& \frac{2}{\epsilon}\mbox{Tr}\{\mathscr{G}(\rho^\prime)\}
+\frac{2 (\epsilon-1)}{\epsilon}\mbox{Tr}\{\Pi\mathscr{G}(\rho^\prime)\}-4\nonumber\\
&=&\frac{4}{\epsilon}\mbox{Tr}\{\rho^\prime\Pi\}
+\frac{(\epsilon-1)}{\epsilon}-4,\nonumber\\
\end{eqnarray}
where, in the last line, we have used the fact that $\Pi$ is the projector for the maximally mixed state $\frac{1}{\sqrt{2}}(\ket{0}+\ket{3})$ and so $\mbox{Tr}\{\mathscr{G}(\rho^\prime)\}=2\,\mbox{Tr}\{\Pi\rho^\prime\}$ and $\mbox{Tr}\{\Pi\mathscr{G}(\rho^\prime)\}=\mbox{Tr}\{\rho^\prime\}/2=1/2$. Hence we obtain the following expression for the variation of the Sinisterness of a Werner state:
\begin{equation}
\delta\mathcal{S}_W = \mathcal{S}_W \frac{4}{\epsilon}\left(\mbox{Tr}\{\Pi\rho^\prime\}-\frac{(3\epsilon+1)}{4}\right).
\label{deltaSW}
\end{equation}



\begin{thebibliography}{99}
	
\bibitem{Schrodinger35} E. Schr\"odinger, ``Discussion of the Probability Relations between Separated Systems," {\em Proceedings of the Cambridge Philosophical Society} {\bf 31}(4), 555-563 (1935).

\bibitem{standardQItext} M. A. Nielsen  and I. L. Chuang, {\em Quantum Computation and Quantum Information} (Cambridge University Press, Cambridge, 2000).

\bibitem{RecentEntanglementReview} 
R. Horodecki, P. Horodecki, M. Horodecki, and K. Horodecki, ``Quantum entanglement,''
{\em Rev. Mod. Phys.} {\bf 81}, 865 (2009).

\bibitem{Wootters98} W. K. Wootters, ``Entanglement of Formation of an Arbitrary State of Two Qubits,'' {\em Phys. Rev. Lett.} {\bf 80}(10), 
2245-2248 (1998).

\bibitem{Werner89} R. F. Werner, ``Quantum States With Einstein-Podolsky-Rosen Correlations Admitting A Hidden-Variable Model,'' {\em Phys. Rev. A} {\bf 40}(8), 4277-4281 (1989).

\bibitem{LangCavesShaji11} M. D. Lang, C. M. Caves and A. Shaji, ``Entropic measures of nonclassical correlations,'' ArXiv e-print 1105.4920 (2011).

\bibitem{QAQJ12} N. Quesada, A. Al-Qasimi and D. F.V. James, ``Quantum properties and dynamics of X states,'' {\em J. Mod. Opt.} {\bf 59}(15), 1322-1329 (2012).

\bibitem{JMKW00} D. F. V. James, P. G. Kwiat, W. J. Munro and A. G. White, ``Measurement of Qubits,'' {\em Phys. Rev. A} {\bf 64}(5), 052312 (2001) (15pp).

\bibitem{Eisert20} J. Eisert, D. Hangleiter, N. Walk et al. ``Quantum certification and benchmarking,'' {\em Nat. Rev. Phys.} {\bf 2}, 382-390 (2020).

\bibitem{Hardy01} L. Hardy, ``Quantum Theory From Five Reasonable Axioms,'' arXiv:quant-ph/0101012 (2001).

\bibitem{Rabi} I. I. Rabi, ``Space quantization in a gyrating magnetic field,'' {\em Phys. Rev.} {\bf 51}, 652-654 (1937).

\bibitem{Cayley}  A. Cayley, ``On the Motion of Rotation of a Solid Body,'' {\em Cambridge Mathematics Journal} {\bf 3}, 224-232 (1843).

\bibitem{Stokes} G. G. Stokes, ``On the composition and resolution of streams of polarized light from different sources,'' {\em Cambridge Philosophical Society Transactions} {\bf 9} 399-416 (1856).

\bibitem{FVH} R. P. Feynman, F. L. Vernon and R. W.  Hellwarth, 
``Geometrical representation of the Schr\"odinger equation for solving maser problems,''
{\em J. Appl. Phys.} {\bf 28}, 49 (1957).

\bibitem{AE} L. Allen and J. H. Eberly, {\em Optical Resonance 
and Two-Level Atoms} (John Wiley, New York, 1975).

\bibitem{DFVJ14} D. F. V. James, ``Quantum Correlations from Classical Coherence Theory?'' {\em Proceedings of the 10th Rochester Conference of Coherence and Quantum Optics}, (J. H. Eberly et al., eds.), 2014.

\bibitem{Peres} A. Peres, {\em Quantum Theory: Concepts and Methods} (Kluwer, Dordrecht, 1995), section 5-3. 

\bibitem{MPSVW} K. Modi, T. Patrek, W. Son, V. Vedral and M. Williamson ``Unified View of Quantum and Classical Correlations,'' {\em Phys. Rev. Lett.} {\bf 104}, 080501 (2010) (4pp).

\bibitem{Fano} U. Fano, ``Pairs of two level systems,'' {\em Rev. Mod. Phys.} {\bf 55}(4), 855-874 (1983).

\bibitem{Englert02} B.-G. Englert and N. Metwally, ``Kinematics of Qubit Pairs,'' in {\em Mathematics of Quantum Computation} (G. Chen and R. K. Brylinski, eds, Chapman and Hall, 2002).

\bibitem{AJMK} J. B. Altepeter, E. R. Jeffrey, M. Medic, and P. Kumar, ``Multiple-Qubit Quantum State Visualization,'' in {\em Conference on Lasers and Electro-Optics/International Quantum Electronics Conference}, OSA Technical Digest (CD) (Optical Society of America, 2009), paper IWC1.

\bibitem{PeresPPT} A. Peres, ``Separability Criterion for Density Matrices,'' {\em Phys. Rev. Lett.} {\bf 77}, 1413?1415  (1996).

\bibitem{SVDref} G. Strang, {\em Introduction to Linear Algebra} (Wellesley-Cambridge Press, 4th ed. 2009), Section 6.7. 


\bibitem{Divincenzo00} D.P. DiVincenzo, B.M. Terhal, and A.V. Thapliyal, ``Optimal decomposition of barely separable states,'' {\em J. Mod. Opt.} {\bf 47}, 377-385 (2000).



\bibitem{negativity} G. Vidal and R. F. Werner, ``Computable measure of entanglement,''  {\em Phys. Rev. A} {\bf 65} 032314 (2002).

\bibitem{OllivierZurek02} H. Ollivier and W. H. Zurek, ``Quantum discord: A measure of the quantumness of correlations,'' {\em Phys. Rev. Lett.} {\bf 88}(1) 017901 (4 pp) (2002).

\bibitem{MorseFesh} P. M. Morse and H. Feshbach, {\em Methods of Theoretical Physics} (McGraw-Hill, New York, 1953), section 7.5, pp.884-886.

\bibitem{Brody} D. C. Brody, ``Biorthogonal quantum mechanics,'' {\em J. Phys. A: Math. Theor.} {\bf 47}, (2014), 035305 (21pp).

\bibitem{SternWalk} M. M. Sternheim and J. F. Walker, ``Non-Hermitian Hamiltonians, Decaying States and Perturbation Theory,'' {\em Phys. Rev. C} {\bf 6}, 114-121 (1972).

\end{thebibliography}
\end{document}